\documentclass[prb,aps,amsmath,amssymb,amsfonts,twocolumn,superscriptaddress,footinbib,longbibliography]{revtex4-1}

\usepackage[colorlinks=true,citecolor=blue,urlcolor=blue,linkcolor=blue,hyperfigures=true]{hyperref}
\usepackage{graphicx}
\usepackage{amsmath,amsfonts}
\usepackage{xcolor}
\usepackage{color}
\usepackage{subcaption}
\usepackage[utf8]{inputenc}

\newcommand{\up}{\uparrow}
\newcommand{\dn}{\downarrow}
\newcommand{\kv}{\ensuremath{\mathbf{k}}}
\newcommand{\qv}{\ensuremath{\mathbf{q}}}

\newcommand{\av}[1]{\ensuremath{\left\langle #1 \right\rangle}}

\renewcommand{\Re}{\operatorname{Re}}

\newcommand{\epsenv}{\varepsilon_{\text{E}}}
\newcommand{\epsmat}{\varepsilon_{\text{M}}}

\usepackage{tikz}
\usetikzlibrary{calc}
\usetikzlibrary{arrows}
\usetikzlibrary{patterns}
\usetikzlibrary{decorations.pathmorphing}
\usetikzlibrary{decorations.pathreplacing}
\usetikzlibrary{decorations.markings}

\definecolor{darkblue}{HTML}{02818A}
\definecolor{darkred}{HTML}{b2182B}
\definecolor{lightred}{HTML}{FDDBC7}
\definecolor{darkgreen}{HTML}{A6D96A}

\begin{document}

\title{Coulomb Engineering of two-dimensional Mott materials}

\author{Erik G. C. P. van Loon}
\affiliation{Institut f{\"u}r Theoretische Physik, Universit{\"a}t Bremen, Otto-Hahn-Allee 1, 28359 Bremen, Germany}
\affiliation{Bremen Center for Computational Materials Science, Universit{\"a}t Bremen, Am Fallturm 1a, 28359 Bremen, Germany}
\affiliation{Mathematical Physics Division, Department of Physics, Lund University, Lund, Sweden}

\author{Malte Sch\"uler}
\affiliation{Institut f{\"u}r Theoretische Physik, Universit{\"a}t Bremen, Otto-Hahn-Allee 1, 28359 Bremen, Germany}
\affiliation{Bremen Center for Computational Materials Science, Universit{\"a}t Bremen, Am Fallturm 1a, 28359 Bremen, Germany}

\author{Daniel Springer}
\affiliation{Institute of Solid State Physics, TU Wien, A-1040 Vienna, Austria}
\affiliation{Institute of Advanced Research in Artificial Intelligence, IARAI, A-1030
Vienna, Austria}

\author{Giorgio Sangiovanni}
\affiliation{Institut f\"ur Theoretische Physik und Astrophysik and W\"urzburg-Dresden Cluster of Excellence ct.qmat, Universit\"at W\"urzburg, 97074 W\"urzburg, Germany}

\author{Jan M. Tomczak}
\affiliation{Institute of Solid State Physics, TU Wien, A-1040 Vienna, Austria}

\author{Tim O. Wehling}
\affiliation{I. Institute of Theoretical Physics, University of Hamburg, D-22607 Hamburg, Germany}
\affiliation{The Hamburg Centre for Ultrafast Imaging, D-22761 Hamburg, Germany}

\begin{abstract}
Two-dimensional materials can be strongly influenced by their surroundings. A dielectric environment screens and reduces the Coulomb interaction between electrons in the two-dimensional material.
Since in Mott materials the Coulomb interaction is responsible for the insulating state, manipulating the dielectric screening provides direct control over Mottness.
Our many-body calculations reveal the spectroscopic fingerprints of such Coulomb engineering: we demonstrate eV-scale changes to the position of the Hubbard bands and show a Coulomb engineered insulator-to-metal transition. 
Based on our proof-of-principle calculations, we discuss the (feasible) conditions under which our scenario of Coulomb engineering of Mott materials can be realized experimentally.
\end{abstract}

\maketitle

\section*{Introduction}

Atomically thin two-dimensional (2d) materials can be influenced by their environment. This idea is utilized in the Coulomb engineering of semiconductors~\cite{Jena07,Rosner16,Raja17,Qiu17}, where the dielectric properties of the environment are used to manipulate the optical and electronic properties such as the carrier mobility~\cite{Jena07,Radisavljevic13}, the band gap~\cite{Rosner16,Borghardt17,Raja17}, quantum hall phenomena~\cite{Papic11} and excitons~\cite{Lin14,Raja17,Gupta17,Druppel17,Florian18,Steinleitner18,Park18,Jia19}. 
This tunability is driven by changes in the screening of the Coulomb interaction. Coulomb engineering is non-invasive in the sense that the semiconducting layer is not changed, only its environment.
As an application, inhomogeneous dielectric environments can be used to produce semiconductor heterojunctions in homogeneous materials~\cite{Rosner16,Utama19}.
In traditional semiconducting 2d materials, the screening causes rigid band shifts~\cite{Waldecker19} and the induced change in the band gap is much smaller than the band gap itself~\cite{Borghardt17,Raja17} (usually 10\%--30\% of the gap). 
In moir\'e correlated electron systems such as twisted bilayer graphene~\cite{Stepanov20,Saito20,Liu21}, environmental screening turned out to modify superconducting critical temperatures as well as transport gaps. 
In metallic systems, on the other hand, intrinsic screening can be so large that environmental screening becomes ineffective~\cite{Schonhoff16}.
Clearly, a detailed analysis of both internal and external screening is required to determine the feasibility of Coulomb engineering for specific applications.

Here, we simulate the Coulomb engineering of Mott insulators and elucidate its spectroscopic fingerprints, which are experimentally accessible via angular resolved photo-emission spectroscopy (ARPES) and scanneling tunneling spectroscopy (STS) experiments.
Since correlations induced by the Coulomb interaction open the gap in the electronic excitation spectrum in Mott insulators~\cite{Imada98}, enviromental screening of the Coulomb interaction in this case holds the potential of not only influencing the band gap to a much larger extent than in semiconductors, but even of closing the gap completely. 

\begin{figure}
   \begin{tikzpicture}[scale=0.6]
  
\path[top color=white,bottom color=white,middle color=lightred] (0,2.25) rectangle (6,5.75) ;
\node at (4.5,4.5) {\Large $\epsmat$} ;

\path[thick,fill=black!20] (0,0) rectangle (6,2);
\node at (3,1) {\Large $\epsenv$} ;
\draw[black,thick] (0,2) -- (6,2) ;

\path[thick,fill=black!20] (0,6) rectangle (6,8);
\draw[black,thick] (0,6) -- (6,6) ;

 \begin{scope}[shift={(2.5,0)}]
 \path[draw=black,inner color=darkred!50, outer color=darkred!0] (0.5,4) .. controls +(-1,2) and +(1,2) .. coordinate (wftop2) (0.5,4) .. controls +(-1,-2) and +(1,-2) .. (0.5,4) ;
 \shadedraw [ball color=darkred,shading=ball] (0.5,4) circle (0.3);  
 \end{scope}

 \draw[thick,dashed] (0.5,3.5) .. controls ++(2.5,-0.8) .. node[above,pos=0.2]{\large $V(r)$} (5.5,3.5) ;
 \draw[thick,dashed] (0.5,3.5) .. controls ++(2.5,-1.8) ..  (5.5,3.5) ;
 \draw[thick,dashed] (0.5,3.5) .. controls ++(2.5,-2.8) ..  (5.5,3.5) ;

 \path[draw=black,inner color=darkred!50, outer color=darkred!0] (0.5,4) .. controls +(-1,2) and +(1,2) .. coordinate (wftop) (0.5,4) .. controls +(-1,-2) and +(1,-2) .. coordinate (wfbottom) (0.5,4) ;
 \shadedraw [ball color=darkred,shading=ball] (0.5,4) circle (0.3);

 \begin{scope}[shift={(5,0)}]
 \path[draw=black,inner color=darkred!50, outer color=darkred!0] (0.5,4) .. controls +(-1,2) and +(1,2) .. (0.5,4) .. controls +(-1,-2) and +(1,-2) .. (0.5,4) ;
 \shadedraw [ball color=darkred,shading=ball] (0.5,4) circle (0.3);  
 \end{scope}

 \draw[<->] (-0.5,2) -- node[left]{$h$} (-0.5,6) ;
 
 \draw[<->] ($(wftop2)+(0,0.25)$) -- node[below]{$a$} ($(wftop)+(0,0.25)$) ;

 \node[anchor=center] at (3,7) {Dielectric environment} ;
 \node[anchor=center,rotate=-90] at (7,4) {Mott layer} ;
\draw [decorate,decoration={brace,amplitude=10pt,mirror},xshift=+4pt,yshift=0pt]
(6,2.5) -- (6,5.5); 
 
 \node at (7,0) {$\phantom{a}$} ;
 
  \end{tikzpicture}
  \includegraphics{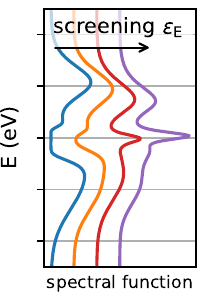}
\caption{
\textbf{Coulomb engineering. a,} A dielectric environment screens the Coulomb interaction $V(r)$ in a monolayer Mott material. \textbf{b,} Environmental screening reduces the gap and can make the Mott material metallic.
}  
\label{fig:sketch}
\end{figure}

\section*{Results}

\subsection{Modelling Coulomb Engineering}

\begin{figure*}[!ht]
  \includegraphics{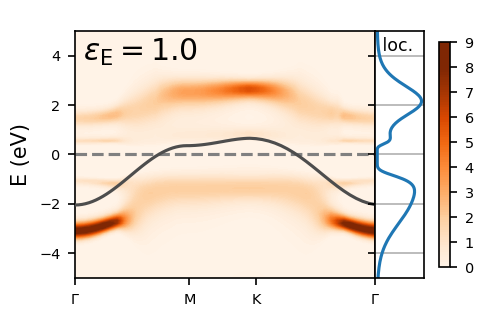} 
  \includegraphics{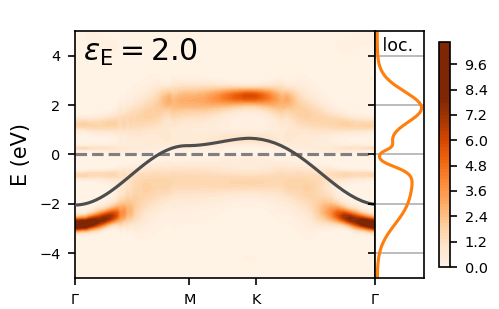} 
  \includegraphics{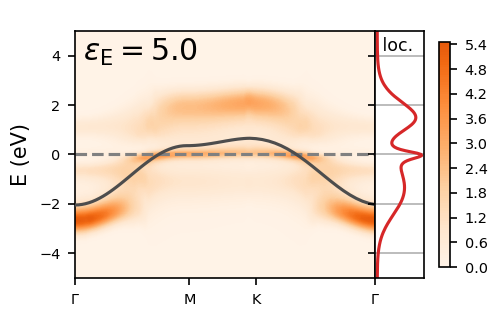} 
  \includegraphics{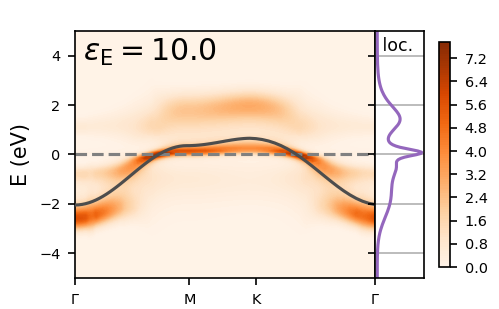} 
 \caption{\textbf{Spectral function $A(k,E)$ of the monolayer depending on the environmental dielectric constant $\epsenv$.} The Fermi level is at $E=0$ (dashed line). The black curve is the non-interacting dispersion. To the right of every momentum-resolved spectral function, the corresponding local part is shown.}
 \label{fig:spectra}
\end{figure*}

Mott insulators are materials that should be metallic according to band theory, but where the electron-electron Coulomb interaction is sufficiently strong to localize the electrons and make the material insulating~\cite{Imada98}.
The ratio of the interaction strength and bandwidth determines whether the potential or the kinetic energy dominates, making the system either insulating or conducting. Coulomb engineering works by changing the interaction strength via the dielectric \emph{environment}, pushing the system towards the conducting phase.

This change from conductor to insulator also dramatically changes the \emph{intrinsic} dielectric properties of the material itself. 
In a good conductor, the internal screening is very efficient and because of this the dielectric environment is less important. 
In the Mott insulator, on the other hand, the electrons are not mobile, and internal screening is inefficient. 
The quantum many-body physics of correlated electrons is the root cause of the reduced internal screening in Mott systems. In two-dimensional systems, such as Mott-insulating ultrathin films, the out-of-plane environment offers an additional pathway to manipulate the screening: This environmental part of Coulomb engineering can be understood on the level of classical electrostatics. Considering a monolayer system, some of the field lines connecting charges in the monolayer leave the material and traverse the surrounding dielectric environment. Thereby, the Coulomb interaction is screened in a peculiar non-local manner: the Coulomb interaction (in the momentum representation) of a monolayer of height $h$ encapsulated in a uniform dielectric environment is given~\cite{Emelyanenko08,Rosner15} by
\begin{align}
   V(q) = \frac{2\pi e^2}{q} \times \frac{1}{\epsmat} \frac{1+x \exp(-qh)}{ 1-x \exp(-qh)}. \label{eq:Vq}
\end{align}
Here $e$ is the electron charge, $q$ is the absolute value of the momentum transfer, $\epsmat$ and $\epsenv$ are the dielectric constants of the material and the environment respectively and $x=(1-\epsenv/\epsmat)/(1+\epsenv/\epsmat)$. 
Since $2\pi e^2/\varepsilon q$ is the usual Coulomb interaction in a two-dimensional material~\cite{Katsnelson12}, the second part of the formula essentially describes the modification of the dielectric function due to the embedding of the monolayer into the dielectric environment. 
$V$ is the effective interaction between the low-energy electrons in the monolayer, so the dielectric constant $\epsmat$ describes all screening in the monolayer \emph{except} for the screening by the low-energy electrons themselves~\cite{Aryasetiawan04}. 

If the monolayer is surrounded by two different materials, a top layer with $x_t$ and a bottom layer with $x_b$, where $x_i = (1-\varepsilon_i/\epsmat)/(1+\varepsilon_i/\epsmat)$ as before, then the effective Coulomb interaction is
\begin{align}
   V(q) = \frac{2\pi e^2}{q} \times \frac{1}{\epsmat} \frac{1+(x_t+x_b) e^{-qh}+x_t x_b e^{-2qh}  }{1-x_t x_b e^{-2qh}}. \label{eq:Vq:onesided}
\end{align}
An important example is a monolayer with vacuum ($\epsilon_{\text{vacuum}}=1$) on one side.
Both one-sided~\cite{Raja17} and two-sided~\cite{Waldecker19} set-ups have been used in experiments and the dielectric model used here accurately describes the experimentally observed Coulomb engineering of conventional semiconductors~\cite{Waldecker19}. 

For demonstrating the possibility and spectroscopic fingerprints of Coulomb engineering in correlated ultra-thin films, we choose specific minimal model parameters which are inspired by transition metal dichalcogenides~\cite{Alhilli72}. We use a triangular lattice with a lattice constant of $a\approx 3.37$ \AA\, a monolayer of height $h=a$ and a background dielectric constant $\epsmat=5$ for the monolayer. Except for Fig.~\ref{fig:onesided}, all results correspond to two-sided screening, equation~\eqref{eq:Vq}. 

In addition to the Coulomb interaction, the kinetic energy is the second ingredient required to describe the Mott insulator. The simplest model involves a single half-filled band of electrons with hopping between neighboring sites.
Combining the potential and kinetic energy results in the extended Hubbard model with Hamiltonian
\begin{align}
 \hat{H} = -t \sum_{\av{ij},\sigma} c^\dagger_{i\sigma} c^{\phantom{\dagger}}_{j\sigma} +\frac{1}{2}\sum_{ij} V(r_i - r_j) n_i n_j. \label{eq:hmlt}
\end{align}
Here $c^\dagger_{i\sigma},c^{\phantom{\dagger}}_{i\sigma}$ are the creation and annihilation operators for an electron with spin $\sigma\in \{\up,\dn\}$ on site $i$ and $n_i = \sum_\sigma c^\dagger_{i\sigma}c^{\phantom{\dagger}}_{i\sigma} $ is the electron density on site $i$.
The Coulomb interaction $V(r)$ is the Fourier transform of equation~\eqref{eq:Vq} and $t$ is the hopping amplitude between neighboring sites $\av{ij}$. We use $t=0.3$ eV, again inspired by transition metal dichalcogenides.


Mott materials, featuring strong correlations, require an advanced many-body treatment~\cite{Georges96,Kotliar06}. 
Here, to understand the Coulomb engineering, we need a consistent treatment of the internal screening in the monolayer, across the insulator-metal transition where the dielectric properties change dramatically. 
To this end, we use state-of-the-art diagrammatic extensions of dynamical mean-field theory~\cite{Rohringer18}, described below. The results in the main text have been obtained with the Dual Boson method~\cite{Rubtsov12,vanLoon14} and were cross-checked with $GW$+DMFT~\cite{Biermann03,Tomczak17,Rohringer18} calculations for a related model in the Appendix.

\subsection{Spectral fingerprints of screening}

Figure~\ref{fig:spectra} shows how the spectral function evolves from $\epsenv=1$ (the freestanding monolayer) to $\epsenv=10$ (encapsulation in bulk Si would correspond to $\epsenv=12$). The colored curve on the right of each graph shows the local density of states, which is the integral of the spectral function over the Brillouin Zone. The screening by the environment leads to substantial changes in the spectral function. Most dramatically, the system changes from an insulator to a metal. 
A comparison of $\epsenv=2.0$ and $\epsenv=5.0$ shows that the gap at the Fermi level disappears and a quasiparticle band emerges at the Fermi level. This is clearly visible both in the spectral function and the local density of states. Experimentally, the latter can be investigated using scanneling tunneling spectroscopy (STS). Below the Fermi level, the momentum-resolved spectral function can be investigated with angular resolved photo-emission spectroscopy (ARPES).

\begin{figure}
  \includegraphics{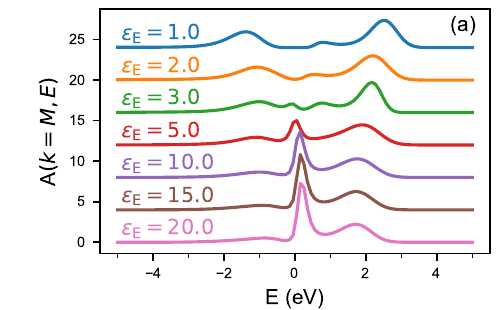} 
  
  \vspace{0.2cm}
  \includegraphics{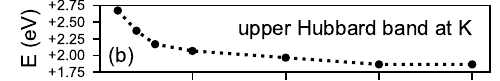}  
  \includegraphics{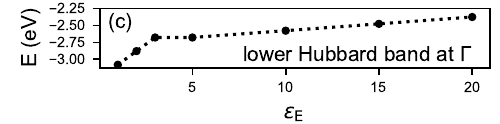}  
  \caption{\textbf{Spectral fingerprints of environmental screening.} \textbf{a,} Energy distribution curve at the M point, i.e., a cross-section of the spectral function at constant momentum. The curves are shifted to increase visibility. \textbf{b, c,} Energy where the upper/lower Hubbard band has maximal spectral weight, for $k=\Gamma$ and $k=\text{K}$, as a function of the environmental dielectric constant $\epsenv$. The electronic temperature is $T=0.1$ eV. }
  \label{fig:EDC}
\end{figure}

Spectral fingerprints of the Coulomb engineering are visible even without crossing the insulator-metal transition: a comparison of $\epsenv=1$ and $\epsenv=2$ shows that the size of the gap within the insulating phase is reduced by the screening. On the metallic side of the transition, the Hubbard side bands are still visible in the spectra and we can trace how their energy changes due to Coulomb engineering. To get a better view, Figure~\ref{fig:EDC}(a) shows the Energy Distribution Curves (EDCs, the cross-sections of the spectral function $A(k,E)$ at fixed momentum) at $k=\mathrm{M}$. Both the lower and upper Hubbard band move towards the Fermi level as $\epsenv$ increases, and for $\epsenv \geq 5$ a quasiparticle peak close to the Fermi level is visible. 
In addition to the changing position of the peaks, the Figure also clearly shows the spectral weight transfer to the quasiparticle peak at the expense of the Hubbard bands.

The Coulomb engineering of Mott insulators leads to eV-sized changes in peak positions, as shown in Figure~\ref{fig:EDC}(b,c).
There, the peak position of the upper and lower Hubbard bands is shown. This energy has been extracted from the EDCs at $k=\Gamma$ and $k=\mathrm{K}$, corresponding to the bottom and top of the band.
The environmental screening changes the position of the Hubbard bands by as much as 0.8 eV. The largest changes occur for small $\epsenv$.

Inside the metallic phase at large $\epsenv$, the effects of environmental screening are also visible in the quasiparticle band, close to the Fermi energy. The spectral weight in this band increases, as was visible in the local spectral function. In addition, the effective electron mass (see ``Methods'') is reduced from $m^\ast/m=3.6$ at $\epsenv=5$ to $m^\ast/m=2.0$ at $\epsenv=20$. 

\subsection{Role of temperature}

\begin{figure}[!ht]
  \includegraphics{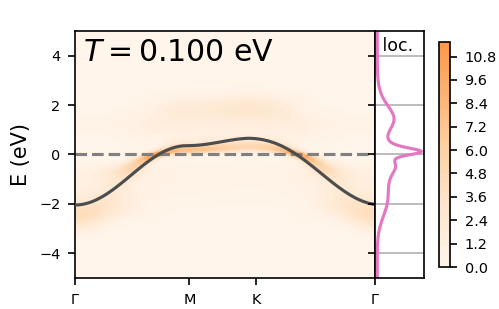} 
  \includegraphics{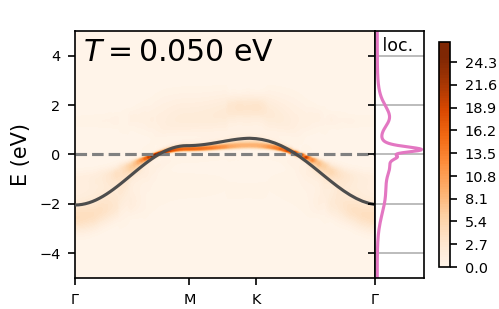} 
  \includegraphics{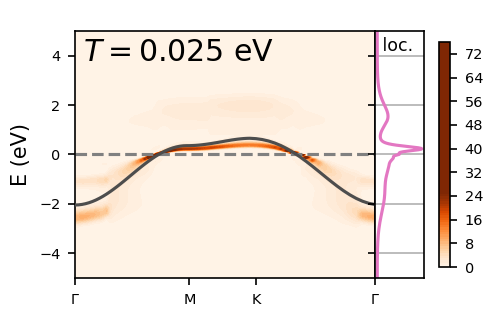} 
 \caption{\textbf{Spectral function $A(k,E)$ of the monolayer depending on the temperature.}  Here, $\epsenv=20$ and the Fermi level is at $E=0$ (dashed line). The black curve is the non-interacting dispersion. On the right of every momentum-resolved spectral function, the corresponding local part is shown.}
 \label{fig:spectra:T}
\end{figure}

The spectral functions of Fig.~\ref{fig:spectra} correspond to an electronic temperature of $T=0.1$ eV (1160 K). 
This temperature is obviously rather high for experiments, however we should stress that we are only simulating the electronic problem, where the energy scale is set by the bandwidth $9t=2.7$ eV. The electronic temperature is small compared to this bandwidth.
On the other hand, the typical energy scale for magnetic phenomena is much lower. 
Our calculations become substantially more difficult and expensive for lower temperatures, limiting the range where we can perform meaningful calculations. 
For $\epsenv=20$, the screening is already quite effective, making the system less correlated, and this reduces the computational cost and allows us to reach lower temperatures.
The spectra at reduced temperatures of $T=0.05$ eV (580 K) and $T=0.025$ eV (290 K, room temperature) are shown in Figure~\ref{fig:spectra:T}. The overall shape of the spectral function does not change substantially in this temperature range, the main difference is that spectral features become sharper at low temperature. 
The bandwidth sets the scale for the temperatures, for a system like magic-angle twisted bilayer graphene, where the bandwidth is two orders of magnitude smaller, our calculations would correspond to meV temperatures (10 K) and all gaps and changes in gaps are also on the meV-scale.

\subsection{Substrate}


\begin{figure}
	\includegraphics{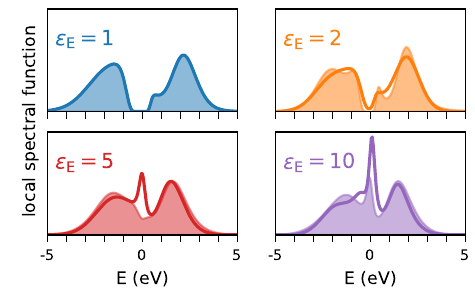}
	\caption{\textbf{Two-sided versus one-sided screening.} The local spectral function for a system surrounded by dielectric constant $\epsenv$ on both sides (lines) or on one side (filled curves). In the latter case, screening is less efficient, resulting in more spectral weight in the Hubbard bands and less in the quasi-particle peak.}
	\label{fig:onesided}
\end{figure}

So far, we have considered the set-up of Fig.~\ref{fig:sketch}, a monolayer surrounded on both sides by a dielectric environment.  
Another experimentally relevant scenario is a Mott monolayer on a dielectric substrate with vacuum on the other side. In that case, the screened Coulomb interaction is given by equation~\eqref{eq:Vq:onesided}. Figure~\ref{fig:onesided} shows the resulting density of states (filled curves) in comparison with two-sided screening (lines). The starting point $\epsenv=1$ is the same, it corresponds to vacuum on both sides. As $\epsenv$ increases, the one-sided screening is less efficient and the resulting spectra retain more Mottish features (smaller quasiparticle peak, Hubbard bands). Still, the overall physics remains the same and Coulomb engineering across the insulator-metal transition is possible. This shows that experiments can use substrate screening instead of both sided screening when it is more convenient. 

\subsection{Not just about $U$}

\begin{table}
\begin{tabular}{r | c c c c }
$\epsenv$ & $V_0$ (eV) & $V_1$ (eV) & $V_2$ (eV) & $V_3$ (eV)  \\
\hline
1 &4.19 & 1.88 & 1.20 & 1.07\\
2 &3.51 & 1.32 & 0.76 & 0.67\\
5 &2.77 & 0.76 & 0.36 & 0.31\\
10 &2.38 & 0.49 & 0.19& 0.16
\end{tabular}
\begin{tikzpicture}[baseline=-0.9]

\node[fill=black,circle] (N0) at (0,0) {\textcolor{white}{$0$}} ;
\node[fill=black!30,circle] at (1,0) {\phantom{1}} ;
\node[fill=black!30,circle] at (0.5,0.86) {$1$} ;
\node[fill=black!30,circle] (N1) at (0.5,-0.86) {$1$} ;
\node[fill=black!15,,circle] (N2) at (1.5,0.86) {$2$} ;
\node[fill=black!15,,circle] at (1.5,-0.86) {$2$} ;
\node[draw,circle] (N3) at (2,0) {$3$} ;

\draw (N0) -- node[sloped,below] {$V_1$} (N1) ;
\draw (N0) -- node[pos=0.85,sloped,below] {$V_2$} (N2) ;
\draw (N0) -- node[pos=0.85,sloped,below] {$V_3$} (N3) ;
\end{tikzpicture}
\caption{\textbf{Screening the Coulomb interaction.} Shown are the on-site, first, second and third neighbour interactions as a function of the dielectric environment. Screening is more effective at larger distances. Note that these four interaction parameters are given here only to illustrate, our model includes interactions on all length scales.  }
\label{tab:interactions}
\end{table}

Internal screening in (quasi-)two-dimensional materials is frequently incomplete, leading to substantial intersite Coulomb interactions~\cite{Wehling11,Hirayama19}, as shown in table~\ref{tab:interactions} for the present model. Environmental screening changes the relative magnitudes of the on-site and intersite Coulomb matrix elements. The influence of intersite interactions on the electronic properties is determined by the spatial extent of charge excitations \cite{Schuler13}: only short-ranged (on-site and nearest-neighbour) interactions are relevant sufficiently deep in the insulating phase, whereas larger interaction lengths are important for the delocalized electrons in a metal. The crux here is that Coulomb engineering pushes the system across the boundary between the two distinct screening regimes, insulator and metal.
Thus, screening in both phases has to be captured appropriately and consistently in a single theoretical description, necessitating advanced theoretical techniques such as the diagrammatic extensions of DMFT used here.


\section*{Discussion}

After this computational demonstration of Coulomb engineering and its spectral fingerprints, we analyse qualitatively how the relevant parameters and scales enter, to guide experimental realizations. 
The effective height $h$ of the monolayer determines the length and momentum scale where the environment becomes important. 
For the limit of large $qh\gg 1$, Equation~\eqref{eq:Vq} reduces to a two-dimensional Coulomb interaction with dielectric constant $\epsmat$. 
In other words, on short length scales the interaction is exclusively determined by the monolayer. The relevant dimensionless parameter is $h/a$, the ratio between the height of the monolayer and the in-plane lattice constant. 
For the Coulomb engineering to be efficient, $h/a$ should not be large. This is confirmed by $GW$+DMFT calculations, as described in the ``Methods'' section.  

The control parameter in our study is $\epsenv$, the dielectric constant of the environment. 
This means that the environment should not influence the monolayer in other ways such as hybridization or (pseudo)doping~\cite{Shao19}, or the bandwidth widening that occurs in, e.g., strontium iridate superlattices~\cite{Kim17}.

The internal dielectric constant $\epsmat$ of the monolayer sets the scale for $\epsenv$, as the expression for $x$ shows. 
Typical values of $\epsenv$ range from 1 for vacuum, via 3.9 for SiO$_2$~\cite{Robertson04} to 12 for Si~\cite{Sze06}.
If $\epsmat$ is large, i.e., if screening inside the monolayer is already very effective, then much larger changes in the dielectric environment are needed to change $V(q)$. For the possibility to turn an insulator into a metal, the monolayer material itself should be an insulator and not too far way from the metallic state. If the goal is only to change the size of the Mott gap then the original material can be deeper inside the Mott phase.

The Hubbard interaction $U$ between electrons on the same site can serve as a useful indicator for the screening possibilities, when we compare a freestanding monolayer ($\epsenv=1$) and the corresponding bulk material ($\epsenv=\epsmat$). 
For the (supposed) Mott insulator CrI$_3$, the Hubbard interaction is reduced from $U = 2.9$ eV for the monolayer to $U = 2.0$ eV for the bulk~\cite{Jang19}.
For the strongly correlated metal SrVO$_3$ there is a reduction from $U=3.7$--$3.95$ eV in the monolayer to $U=3.3$ eV in the bulk~\cite{Zhong15}.
These substantial reductions of the Coulomb interaction are achieved by replacing vacuum with other Mott layers, meaning that a more effective dielectric environment will reduce the Coulomb interaction even more.

Meanwhile, the interlayer spacing and in-plane lattice constants in CrI$_3$ are both approximately 7 \AA\, so that $h/a\approx 1$~\cite{Sivadas18}.
Layered cuprate materials, another material family known for strong correlation physics, is worse in this respect, with typical values~\cite{Murphy87} of $a\approx 4$ \AA\ and $h \approx 12$ \AA, so $h/a=4$. While it is possible to exfoliate cuprate single layers~\cite{Novoselov05b}, their thickness compared to the lattice constant renders Coulomb engineering more challenging.
For transition metal dichalcogenides, $h/a <2$ can be achieved~\cite{Alhilli72}, which is better than the cuprates but not as good as CrI$_3$.

It is even possible to reach $h<a$ when $a$ is the size of an emergent superlattice which can be much larger than the interatomic distance of the underlying lattice. One example is twisted bilayer graphene~\cite{Bistritzer11,Cao18a,Cao18b} with a moir\'e superlattice hundreds of times larger than the graphene unit cell so that $h/a$ can become genuinely small. We should note that $\epsmat$ is quite large in twisted bilayer graphene~\cite{Pizarro19} and that the insulating state is not of the idealized Mott-Hubbard type but likely involves spontaneous symmetry breaking in the valley and spin degrees of freedom~\cite{Lu19,Sharpe19,Jiang19,Serlin20}. 
Correspondingly, the interpretation of first possible experimental reports~\cite{Stepanov20,Saito20,Liu21} of Coulomb engineering related effects in twisted bilayer graphene is intricate~\cite{Pizarro19,Goodwin19}. 

A clearer situation appears for certain charge density wave phases in 2d materials:
1T-TaS$_2$~\cite{Tosatti76,Fazekas79,Fazekas80}, NbSe$_2$ and TaSe$_2$ have so-called commensurate charge density waves with a star-of-David lattice reconstruction, the emergent scale is $\sqrt{13}$ times the original lattice constant.
The star-of-David reconstructed TaSe$_2$ monolayer has a Mott gap of approximately 0.1 eV~\cite{Chen20}, and lends itself as a natural candidate for the exploration of Coulomb engineering of Mott materials.

A final important point for experimental realizations is the specific shape and extent of the correlated orbitals. This is of relevance for the interaction $V(r)$ when $r$ is small. Our modelling assumes that the charge density of the electronic orbitals is homogeneous on the scale $a$ of the unit cell. A Wannier function with radius $r_\mathrm{WF} < a$ will have an increased local interaction, whereas the interaction between electrons in different unit cells is less affected by the Wannier radius. As an example, an Ohno fit~\cite{Ohno64}, $V(r)=V_0/\sqrt{1+r^2/r_\mathrm{WF}^2}$, to the on-site and nearest-neighbor interactions in cuprates~\cite{Hirayama19} gives $a/r_\mathrm{WF}\approx 4.5$, whereas the same Ohno model in graphene~\cite{Schuler13} gives $a/r_\mathrm{WF}\approx 1.5$. Since the dielectric screening is more efficient on longer length scales, Coulomb engineering is more favourable for materials with a larger Wannier radius. 

\section{Conclusion}

We have demonstrated that the dielectric environment can be used to control Mott insulating layered materials and that Coulomb engineering across the insulator-metal transition is possible. Our calculations show the spectroscopic fingerprints of Coulomb engineering, namely eV-scale movement of the Hubbard bands and the appearance of a quasiparticle band as the system turns metallic. 
Based on our modelling, we can identify necessary conditions for Coulomb engineering. To be effective, the dielectric environment should be close to the monolayer, while also avoiding other monolayer-environment couplings like pseudodoping and hybridization.
Furthermore, the monolayer itself should not screen too strongly. A good indicator is if the Hubbard parameter $U$ is much larger for the monolayer than for the corresponding bulk material.
Our results open a perspective for the fabrication of heterostructures by the application of dielectric covering on parts of a Mott monolayer. In this way, the environment can create local phase transitions in correlated materials: metallic paths in an otherwise insulating layer or the control of unconventional superconducting phases via the reduction of the effective interaction. The sharpness of these dielectrically controlled heterostructures is controlled by the electronic Green's function~\cite{Rosner16}, so in Mott insulators where the electrons are localized, heterostructures can be atomically sharp.






\section*{Methods}

\subsection*{Dual Boson}

The system is studied in the grand canonical ensemble with density fixed to half-filling.
For our computations, we use the Dual Boson method~\cite{Rubtsov12} with the implementation described in Ref.~\onlinecite{vanLoon14}.
It consists of a self-consistency cycle to determine the optimal Anderson Impurity Model (AIM). In this cycle, we include vertex corrections to the susceptibility in the ladder approach to ensure that the susceptibility satisfies charge conservation at small momenta. This is important when dealing with long-ranged $V(q)$.
When self-consistency has been reached, we calculate the spatial self-energy $\tilde{\Sigma}(k,\nu)$ in the last iteration using the second-order diagram of Figure 2b from~\onlinecite{vanLoon14}, which together with the AIM self-energy is used to determine the Green's function $G$ and finally the spectral function $A(k,E)$.
The diagrammatic calculations are done on a $64\times 64$ periodic lattice, sufficient to rule out electronic finite-size effects, especially since the electrons are rather localized in the regime studied here. We study the non-magnetic insulator-metal transition, we do not allow for magnetic order in our calculations. 
The Anderson Impurity Model is solved using the ALPS/w2dynamics (for $GW$+DMFT, see below) CT-HYB solvers~\cite{ALPS2,Hafermann13,Hafermann14,Wallerberger19}. 
All our calculations are done in Matsubara space, the analytical continuation to real energies is done to determine $A(k,E)$. For this, we use OmegaMaxEnt\cite{Bergeron16}. To verify the reliability, we have also performed stochastic continuation using Spektra (\url{https://spektra.app})~\cite{Ghanem17,Ghanem20} for a representative subset of our calculations.

\subsection*{Effective mass}

In the metallic phase, the effective electronic mass is a useful indicator of the strength of correlations. 
The effective mass renormalization consists of two parts~\cite{Tomczak14,Schafer15b}, a dynamical contribution $\left(1-\frac{\partial \Re \Sigma(\kv,\omega)}{\partial \omega}\right)^{-1}$ and a static contribution. The dynamical contribution decreases as the Coulomb interaction is reduced by screening, but the spatial contribution actually increases due to Fock-like band widening. The latter effect turns out to be smaller, leading to an overall decrease of the effective mass upon increased screening. The values of the effective mass given in the Results section were determined on the $\Gamma$-M high symmetry line, at the Fermi surface crossing. In our calculations, the effective mass depends only weakly on momentum.

\subsection*{$GW$+DMFT}

\begin{figure}[t]
  \includegraphics{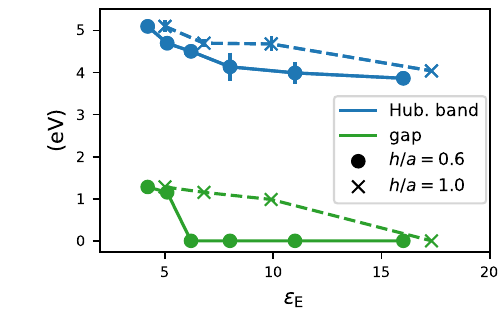}
  \caption{\textbf{Influence of $h/a$.} In blue, the energy where the Hubbard band in the square lattice model has maximal spectral weight as a function of the environmental dielectric constant $\epsenv$. This maximum is determined for $k=\Gamma$ and $k=$M, the symbols show the average and the error bars the difference. In green, the magnitude of the gap.}
  \label{fig:DMFTGW:gaps}
\end{figure}

\begin{figure}[t]
   \includegraphics{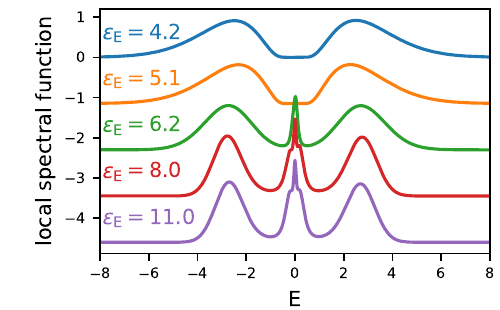}
   \caption{
\textbf{Local spectral function of the Hubbard model on the square lattice.} $GW$+DMFT results are shown as a function of the environmental screening, with $h/a\!=\!0.6$ constant. 
}
   \label{fig:DMFTGW:DOS}
 \end{figure}
 
To illustrate that our findings are representative of Mott monolayers in general, we have also performed $GW$+DMFT~\cite{Biermann03} calculations for a square lattice system, based on the implementation of Ref.~\cite{Schafer15b,Tomczak17}. 
The non-local contributions of the self-energy, calculated in a single-shot GW calculation, are added to the local DMFT self-energy~\cite{Wallerberger19}.

The $GW$+DMFT treatment of the interaction in this model was restricted to local ($U$) and nearest-neighbor ($V_\text{nn}$) terms. To make a comparison with the dielectric model of equation~\eqref{eq:Vq}, we map the $(\epsenv,\epsmat,h/a)$ model onto the two-parameter effective interaction $\hat{V}(\qv)=U + 2V_\text{nn} [\cos(q_x)+\cos(q_y)]$ via two constraints: 
\begin{align}
 \int_{\text{BZ}} V(\qv) d\qv =& \int_{\text{BZ}} \hat{V}(\qv) d\qv, \\
 V(\pi,\pi) =& \hat{V}(\pi,\pi).
\end{align}
The first constraint ensures that the overall magnitude of the interaction is the same in the effective model, which is important for the Mott physics, the second constraint that the potential energy cost of checkerboard charge-density waves is equal for both interactions. The latter is important to ensure that the effective model is not biased towards the CDW phase of the square lattice extended Hubbard model~\cite{Hansmann13}.
For the material parameters, we used $a=2.27$ \AA, $\epsmat=2.9$ and both $h/a=0.6$ and $h/a=1$.
The hopping was set to $t=0.5$ eV, the temperature to $T=0.025$ eV (290 K, room temperature).

The numerical advantage of the $GW$+DMFT approach compared to the Dual Boson method is that it does not involve so-called vertex corrections, which makes it substantially cheaper and allows us to explore larger parts of phase space even at low temperatures. Here, we use it to study the effect of $h/a$. At the same time, the neglect of vertex corrections is an approximation, which is potentially problematic for collective excitations on longer length scales. This is why we cut off the interaction after the nearest-neighbors in this model. The square lattice at half-filling has particle-hole symmetry, for the present investigation this, for example, means that the lower and upper Hubbard band should appear at the same (absolute) energy, which is used to estimate the error bars in figure~\ref{fig:DMFTGW:gaps}, coming from uncertainty in the analytical continuation.

The results are shown in figures \ref{fig:DMFTGW:gaps} and \ref{fig:DMFTGW:DOS}. As in the Dual Boson results, the $\epsenv$ induced insulator-metal transition is clearly visible. Here, this transition occurs around $\epsenv=5$ for $h/a=0.6$ and $\epsenv=10$ for $h/a=1$. 
As $\epsenv$ increases, the Hubbard bands move towards the Fermi surface. 

\acknowledgments
The authors thank P. Hofmann for useful discussions and K. Ghanem for the help with the stochastic continuation. 
The authors acknowledge the North-German Supercomputing Alliance (HLRN) for providing computing resources via project number hbp00047 that have contributed to the research results reported in this paper.
D. S. and J. M. T. acknowledge support by the Austrian Science Fund (FWF) through project 'LinReTraCe' P 30213-N36. Some calculations were performed on the Vienna Scientific Cluster (VSC). 
G. S. acknowledges financial support from the DFG through the W\"urzburg-Dresden Cluster of Excellence on Complexity and Topology in Quantum Matter -- \textit{ct.qmat} (EXC 2147, project-id 39085490). T.W. acknowledges support from the DFG through QUAST (FOR 5249, No. 449872909) and via the Cluster of Excellence ``CUI: Advanced Imaging of Matter'' (EXC 2056, No. 390715994).

\section*{Author contributions}

EvL performed the Dual Boson calculations, MS performed the analytical continuations, DS and JMT performed the $GW$+DMFT calculations. All authors contributed to the interpretation of the results and to the writing of the manuscript.

\bibliography{references}

\end{document}